\documentclass[prb,twocolumn,showpacs,preprintnumbers,amsmath,amssymb]{revtex4}
\usepackage{dcolumn}
\usepackage{bm}
\usepackage{amsmath,graphicx}
\usepackage{color}
\usepackage{times}

\def\d{{\rm d}}

\def\e{{\rm e}}
\def\im{{\rm i}}

\def\fu{C$_{60}$ }

\begin{document}
\title{Tomonaga-Luttinger liquid theory for metallic fullurene polymers}

\author{Hideo Yoshioka}
\affiliation{Department of Physics, Nara Women's University, Nara 630-8506, Japan}

\author{Hiroyuki Shima}
\affiliation{Department of Environmental Sciences, University of Yamanashi, 4-4-37, Takeda, Kofu, Yamanashi 400-8510, Japan}
 
\author{Yusuke Noda}
\affiliation{RIKEN Innovation Center, Nakamura Laboratory, 2-1 Hirosawa, Wako, Saitama 351-0198, Japan\\}
\affiliation{Department of Physics, Yokohama National University, 79-5, Tokiwadai, Yokohama 240-8501, Japan}

\author{Shota Ono and Kaoru Ohno}

\affiliation{Department of Physics, Yokohama National University, 79-5, Tokiwadai, Yokohama 240-8501, Japan}

\date{\today}

%
%

\begin{abstract}
We investigate the low energy behavior of local density of states 
in metallic C$_{60}$ polymers theoretically. 
The multichannel bosonization method is applied 
to electronic band structures evaluated from first principles calculation, 
by which the effects of electronic correlation and nanoscale corrugation in the atomic configuration are fully taken into account.
We obtain a closed-form expression for the power law anomalies 
in the local density of states,
which successfully describes the experimental observation
on the \fu polymers in a quantitative manner.
An important implication from the closed-form solution 
is the existence of an experimentally unobserved crossover at nearly a hundred milli-electron volts,
beyond which the power law exponent of the \fu polymers should change significantly.
\color{black}
\end{abstract}





\maketitle
\section{Introduction}
Tomonaga-Luttinger liquid (TLL) state is a remarkable quantum state
unique to one-dimensional (1D) correlated electron systems \cite{Tomonaga1950,Luttinger1963}. 
One of the characteristics is the spin-charge separation,
by which  excitations of spin and charge travel independently through the system 
at different speeds \cite{Haldane1980}.
Another notable manifestation of the TLL state is a series of power-law anomalies
observed in the correlation functions of physical quantities, 
the single-particle density of states (DOS), and 
the momentum distribution.
For instance, the DOS $D(\omega,T)$ near the Fermi energy $E_\mathrm{F}$ 
obeys $D(\omega, 0) \propto |\omega|^\lambda$ and 
$D(0, T) \propto T^\lambda$,  
where $\omega$ is the energy measured from $E_\mathrm{F}$ and $T$ is the
absolute temperature.
The value of the exponent $\lambda$ depends on 
the mutual interaction between electrons \cite{Voit1995},
and can be controlled by the external field \cite{Bellucci2006,Klanjsek2008,Shima2010}.

Experimental realization of TLL states has been achieved thus far
in a wide range of materials with highly anisotropic conductivity
\cite{Aleshin2004,Auslaender2005,Hager2005,Venkataraman2006,BJKim2006,Yuen2009,Jompol2009}.
A typical example is metallic carbon nanotube having a shape of rolling up a graphene sheet
\cite{Bockrath1999,Yao1999,Bachtold2001,Ishii2003,Tombros2006,Danilchenko2010}.
It was reported that metallic carbon nanotubes show power laws
in their differential conductance $d I/d V \propto |V|^\lambda$ measured at
high bias voltage ($eV \gg T$) and in their temperature-dependent conductance 
$G \propto T^\lambda$.
More interestingly, the exponent is highly sensitive to boundary effects;
the power-law behavior of a finite-sized TLL with an open boundary
differs from that of an infinite-sized TLL.\cite{Fabrizio1995,Eggert1996,Mattsson1997}
In the conductance measurements, in fact,
the exponent obtained under the edge-contact condition was much larger than
that obtained under the bulk-contact condition.\cite{Bockrath1999}
These transport anomalies in the carbon nanotubes result from 
the location dependence of local density of states (LDOS) \cite{Yoshioka2003},
as has been directly observed through
photoemission spectra (PES) \cite{Ishii2003}.

In the last years, a novel class of nanocarbon, called the 1D \fu polymer,
was proved to exhibit the TLL behavior.\cite{Shima2009,Onoe2012}
This 1D nanomaterial is synthesized by 3 keV electron beam irradiation to two-dimensional \fu films.\cite{Onoe2003}
The irradiation induces sequential in-plane rotations of carbon-carbon (C-C) bonds 
followed by polymerization of adjacent \fu molecules,
resulting in a long, thin, peanut-shaped nanocarbon cylinder.
From the first synthesis, many unexpected features have been confirmed 
in low-energy excitations of the 1D \fu polymers \cite{Onoe2004,Onoe2007,Toda2008,Takashima2010,Shota2011,Ryuzaki2014,Shota2014},
which are largely different from those of straight-shaped counterparts ({\it i.e.,} carbon nanotubes).
Such the difference was also observed in the power-law exponent decribing the TLL states.
A recent experiment of {\it in situ} ultraviolet PES indicated that
the exponent for the 1D \fu polymer is significantly larger
than that of metallic carbon nanotubes.\cite{Onoe2012}
The increase in the exponent is potentially attributed to the periodic insertion
of non-hexagonal carbon rings into a rolled-up graphene sheet,
which causes periodic modulation in the tubular radius along the tubular axis\cite{Shima2009}.
Nevertheless, no theoretical attempt has been succeeded to describe
the \fu polymer's exponent in a quantitative manner.
To resolve the problem, 
it is dispensable to apply a multichannel-based TLL theory\cite{Yoshioka2011}
incorporated with precise information of energetically stable atomic structure of 
the 1D \fu polymers.\cite{Noda2015}

In the present work, we extend a conventional bosonization theory
to be applicable to 1D systems endowed with multichannel electronic conduction bands.
The theory we have developed yields a closed form solution for the LDOS
both under the edge-contact and bulk-contact conditions,
thus allowing us to describe the TLL behavior of the 1D \fu polymers in a quantitative manner.
We demonstrate that the power law exponent deduced from the theory agrees with
the existing experimental observation.
The theory also leads us to a prediction that
the \fu polymers should exhibit a crossover in the power law of the LDOS
at a characteristic energy or temperature, beyond which the value of the exponent shifts significantly.

\section{First principles calculation}

\begin{figure}[ttt]
\includegraphics[width=8.0cm,clip]{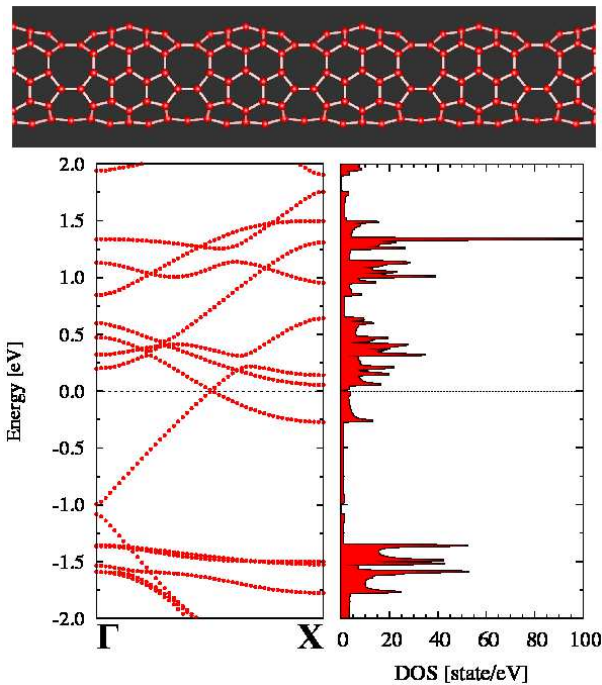}
\caption{
(Top) Atomic configuration of the FP6L model.
The waist part is occupied by a combination of heptagonal and octagonal rings of carbon atoms.
(Bottom) The dispersion relation and 
the electronic density of states of the FP6L model.
}
\label{fig:FP6L}
\end{figure}
\begin{figure}[ttt]
\includegraphics[width=8.0cm,clip]{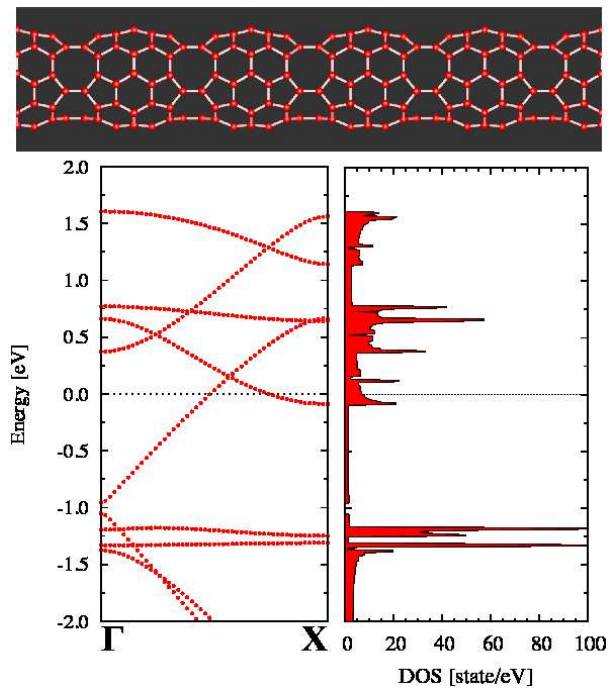}
\caption{
(Top) Atomic configuration of the FP5N model.
The waist part is occupied by five octagonal rings of carbon atoms.
(Bottom) The dispersion relation and 
the electronic density of states of the FP5N model.}
\label{fig:FP5N}
\end{figure}

\subsection{Stable atomistic configuration}

We performed first principles calculations to obtain 
the energetically stable atomic models for the 1D \fu polymer \cite{Noda2011,Noda2014,Noda2015}.
Our simulations are initiated by setting the T3 model \cite{GWang2005},
a semiconducting 1D \fu polymer with energy gap $\sim$ 1.17 eV.
Inducing a sequence of in-plane rotations at specific parts of 
C-C bonds\cite{Stone1986,SHan2004} to the T3 model,
we have created more than fifty kinds of metallic and semiconducting 1D \fu polymers 
with different atomic configurations.
All the first principles calculations were performed
using Vienna {\it ab initio} simulation package\cite{Kresse1996} 
based on a density functional theory \cite{Hohenberg1964}.
Using the $8\times 1\times 1$ Monkhorst-Pack $k$-point grid,
we optimized all the structures until atomic forces fall below 10$^{-4}$ eV/\AA\; on each atom.

Through the analysis, we found that
the total energy and the stability of the resulting models
depend on the number of non-hexagonal rings ({\it i.e.,} pentagon, heptagon, and octagon)
embedded and their relative positions.
Among many candidates, we focus on the two most stable models labeled by\cite{Noda2015} FP6L and FP5N;
see the top panels in Figs.~\ref{fig:FP6L} and \ref{fig:FP5N}.
The total energy of the FP6L model per unit cell is 1.421 eV lower than the original T3 model,
and that of FP5N is 2.188 eV lower than T3.
As for the FP6L model, it is seen from Fig.~\ref{fig:FP6L} that
the waist part is composed of a combination of heptagonal and octagonal rings of carbon atoms.
On the other hand, the waist part of the FP5N model is occupied five octagons.
This slight difference in the lattice symmetry between the two models
causes a drastic change in the Fermi level crossing of electronic energy bands,
as explained in the next subsection.

\begin{table}[bbb]
\caption{Fermi velocities of 1D \fu polymers estimated from the band structures I and II.}
\begin{tabular}{|c|c|c|} \hline
          & Band structure I        & Band structure II       \\ \hline
$v_1$     & \;\; $6.809 \times 10^5$ m/s\;\; & \;\;$7.322 \times 10^5$ m/s\;\; \\
$v_2$     & \;\; $4.237 \times 10^5$ m/s\;\; & \;\;$2.639 \times 10^5$ m/s\;\; \\
\;\;$v_2/v_1$\;\; & \;\; $0.6223$               \;\; & \;\;$0.3604$               \;\; \\ \hline
\end{tabular}
\label{table_fermi}
\end{table}

\subsection{Energy dispersion near the Fermi level}

The electronic energy dispersion and DOS derived from the two models
are shown in the lower panels in Figs.\ref{fig:FP6L} (FP6L) and \ref{fig:FP5N} (FP5N).\cite{Noda2015}
In both figures, the Fermi energy is indicated by the dashed horizontal lines. 
Because of the existence of dispersion curves
that cross the Fermi level,
metallic conduction is expected for the both models. 
In the following, the energy dispersion derived from the FP6L model is called as the band structure I,
and that from FP5N is called the band structure II.

As seen from the lower panel in Fig.~\ref{fig:FP6L},
the band structure I possesses only one Fermi-level-crossing point,
just on which two dispersion curves with Fermi velocities (positive or negative)
come across.
In addition, the band structure I comprises only nondegenerate dispersion curves.
These features in the dispersion profile looks similar to 
those observed in metallic carbon nanotubes.\cite{Kanamitsu2002}
There are, nevertheless, two important differences in the dispersion profile
between the band structure I and the carbon nanotubes as explained below.

The first difference to be remarked is 
the asymmetry in the slopes of the two Fermi-level-crossing curves in the band structure I.
Namely, the X-shaped band crossing at the Fermi level
is a little slanted, in contrast with the fully symmetric band crossing 
in the metallic carbon nanotubes.
It is interesting to note that 
an asymmetry similar to that in the present system
has been realized at the asymmetric Dirac cone in the two dimensional
organic materials, $\alpha$-(ET)$_2$I$_3$.\cite{Kajita2014}

The second to be mentioned 
is a disagreement in the absolute value of the Fermi velocity
between the band structure I and the carbon nanotubes.
As will be shown quantitatively,
the absolute values of the Fermi velocities evaluated from the band structure I
are much smaller than that of graphene sheets $v_0 \simeq 8.0 \times 10^5 \mathrm{m/s}$.
This ``slowness" of the present system
causes a considerable enhancement in the power-law exponent $\lambda$
compared with those of the carbon nanotubes; we will revisit details on the issue later.

Similar discussion as above holds for the band structure II
that has been presented in the lower panel of Fig.~\ref{fig:FP5N}.
An interesting disparity from the case of I is that
the band structure II contains doubly degenerate dispersion curves
because of high degree of symmetry in their atomistic configuration.
Accordingly, the band structure II involves
three crossing points at the Fermi level, 
since the dispersion curve with the negative Fermi velocity is doubly degenerate. 
Due to the degeneracy, the interband crossing point of these three bands is repelled from the Fermi level.

\begin{figure}[ttt]
\includegraphics[width=9.5cm]{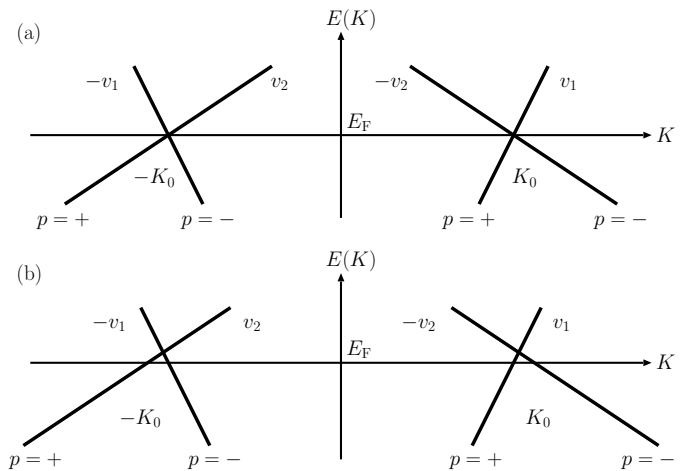}
\caption{
Sketch of linear dispersions close to Fermi energy $E_\mathrm{F}$.
(a) The band structure I derived from the FP6L model, in which
two bands cross at $\pm K_0$. The label $p = +/-$ indicates the right/left moving state. 
Fermi velocities are denoted as $v_1$ and $v_2$ satisfying $v_1 > v_2$.
(b) The band structure II derived from the FP5N model, in which
the interband crossing points deviate from the Fermi level.
Note that the band with the Fermi velocity $\pm v_2$ is doubly degenerate. 
}
\label{fig:band}
\end{figure}

\subsection{Simplified view of Fermi level crossing}

The energy dispersions close to Fermi energy 
are schematically given in Figs.~\ref{fig:band} (a) and (b)
for the case of I and II, respectively. 
In the band structure I, 
two bands with the linear dispersion cross the Fermi energy at
$\pm K_0$, which correspond to the so called K and K' point of the
graphene. 
We denote the Fermi velocities of the two bands $v_1$ and $v_2$,
respectively, and use the label
$p = +(-)$ to express the right (left) moving states. 
Differing from the case I,
there exist three bands crossing the Fermi energy in the band
structure II.  
The band with the larger velocity $v_1$ is not degenerate, whereas
those with the smaller one $v_2$ are doubly degenerate.
Because of the presence of three (not two) bands across the Fermi level,
the crossing point $\pm K_0$ is allowed to deviate from the Fermi wavenumber.
The actual values of Fermi velocity in the models I and II are summarized in Table \ref{table_fermi}.

\section{Bosonization approach}

\subsection{Hamiltonian}

We now formulate an effective Hamiltonian that describes single-particle excitations.
The kinetic term of the Hamiltonian, $\mathcal{H}_\mathrm{k}$, reads
\begin{align}
 \mathcal{H}_\mathrm{k} &= \sum_{k} \sum_{i=1}^{N_b} \sum_{p=\pm}
 \sum_{s=\uparrow, \downarrow} p v_i k c^{\dagger}_{i,p,s}(k)
 c_{i,p,s}(k), 
\end{align} 
where $c_{i,p,s}(k)$ denotes the annihilation operator of an electron
with the $i$-th band, branch $p$, spin $s$ and wavenumber $k$.
The wavenumber $k$ is
measured from the Fermi wavenumber of the each band. 
The number $N_b$ is equal to 2 and 3 in the band structures I and II,
respectively. 
Note that $v_3 = v_2$ in the case II.

We take into account of the long range Coulomb interaction as a mutual
interaction. 
The term with the largest matrix element is written as 
\begin{align}
 \mathcal{H}_\mathrm{int} &= \frac{\bar{V}(0)}{2} \int dx \rho(x)^2. 
\end{align}
Here $\rho(x)$ is the slowly varying component of charge density
written by $\rho(x) = : \sum_{i,p,s} \psi^\dagger_{i,p,s}(x)
\psi_{i,p,s}(x):$ with 
$\psi_{i,p,s}(x) = 1/\sqrt{L} \sum_k \mathrm{e}^{\im k x}
c_{i,p,s}(k)$ and $L$ being the length of the system.  
The matrix element of the forward scattering $\bar{V}(0)$ is given by 
$\bar{V}(0) = 2 (e^2/\kappa) \ln R_s/R$.  
Here $\kappa$ is the
dielectric constant, $R$ is the average radius of the 1D \fu polymer and 
$R_s$ is the large distance cut-off of the Coulomb interaction 
which corresponds to the length of the polymer. 
The matrix element is insensitive to choices of such length scales 
due to its logarithmic dependences.    
We discard the terms with the matrix elements on the order of
$a/R$ with $a$ being the length scale of the order of C-C bonds
 which may give rise to the tiny energy gaps 
as is seen in carbon nanotubes.\cite{Egger1997,Egger1998,Yoshioka1999,Odintsov1999}
Therefore, the treatment in the present study is effective for the case where the
temperature $T$ or the energy $\omega$ measured from the Fermi energy 
is much larger than such gaps.

In our previous study, we developed the TLL theory
of multi-channel one-dimensional systems where the Fermi
velocity is different from each other.\cite{Yoshioka2011} 
Here, the rigorous expression of the LDOS was
derived for the bulk and end position. 
In the following,    
we apply the theory to the band structures I and II of 
the one-dimensional \fu polymer obtained by the first
principles calculation. 

For this purpose, for the band structure I and II, 
we introduce the index $\nu = 1, 2, 3, 4$ and $\nu = 1, 2, 3, 4, 5, 6$,
respectively. 
In the case I, 
$\nu = 1 (3)$ corresponds to the band with the Fermi velocity $v_1$ and the
up (down) spin, whereas 
$\nu = 2 (4)$ corresponds to that with the Fermi velocity $v_2$ and the
up (down) spin.   
On the other hand, for the case II,  
$\nu = 1 (4)$ expresses the band with $v_1$ and the up (down) spin, 
whereas $\nu = 2,3 (5,6)$ are those with $v_2$ and the up (down) spin.

Then the Hamiltonian is written as 
\begin{align}
 \mathcal{H}_\mathrm{k} &= \sum_{\nu=1}^N \sum_{p=\pm} \sum_{k}
 p v_{\mathrm{F}\nu} k c^{\dagger}_{p,\nu}(k) c_{p,\nu}(k), \\
 \mathcal{H}_\mathrm{int} &= \frac{\bar{V}(0)}{2L}  
\sum_{\nu,\nu'=1}^N
\sum_{p,p'=\pm} \sum_q  \rho_{p,\nu}(q) \rho_{p',\nu'}(-q), 
\end{align} 
with $N=4$ for the band structure I or $N=6$ for II. 
Here, $v_{\mathrm{F}1} = v_{\mathrm{F}3} = v_1$ and  
$v_{\mathrm{F}2} = v_{\mathrm{F}4} = v_2$ in I, whereas 
$v_{\mathrm{F}1} = v_{\mathrm{F}4} = v_1$ and 
$v_{\mathrm{F}2} = v_{\mathrm{F}3} = v_{\mathrm{F}5} = v_{\mathrm{F}6} =
v_2$ in II.  
The quantity $\rho_{p,\nu}(q)$ is defined by 
$\rho_{p,\nu}(q) = :\sum_k c^\dagger_{p,\nu}(k+q) c_{p,\nu}(k):$.   

According to Ref.~\onlinecite{Yoshioka2011}, 
the above Hamiltonian $\mathcal{H} = \mathcal{H}_\mathrm{k} +
\mathcal{H}_\mathrm{int}$ is expressed by the phase variables $\Theta_j$ and
$\Phi_j$ as  
\begin{align}
 \mathcal{H} &= \frac12 \sum_{j=1}^N \int \d x
\left\{ \Xi_j^2 + \tilde{v}_j^2 (\partial_x \Theta_j)^2 \right\},
\label{eqn:a4}
\end{align}  
where these variables satisfy the relation, $[\Theta_j(x),\Phi_{j'}(x')] = \im \delta_{jj'} 2 \pi
\theta(x-x')$ and $\Xi_j(x) = - \partial_x \Phi_j(x)/(2 \pi)$, with 
$\theta(x)$ being the conventional step function.  
The velocities of the excitation $\tilde{v}_j$ is determined by the following
eigenvalue equation,
\begin{align}
 (\tilde{v}_j^2 K - V) \bm{X}_j = 0, 
\label{eq_shima25}
\end{align} 
where the matrices $K$ and $V$ are given as follows,
\begin{align}
 (K^{-1})_{\nu,\nu'} &= 2 \pi v_{\mathrm{F}\nu} \delta_{\nu,\nu'}, \\
 V_{\nu\nu'} &= \frac{v_{\mathrm{F}\nu}}{2 \pi} + \frac{\bar{V}(0)}{2
 \pi^2} \; =  \frac{v_{\mathrm{F}\nu}}{2 \pi} + g. 
\end{align}
The eigenvector $\bm{X}_i$ is normarized as $(\bm{X}_i,K\bm{X}_j) =
\delta_{ij}$. 

The electron operator, $\psi_{p,\nu}(x) = (1/\sqrt{L}) \sum_k
\e^{\im k x} c_{p,\nu} (k)$, is also written in terms of phase variables
as  
\begin{align}
 \psi_{p,\nu}(x) &= \frac{\eta_{\nu}}{\sqrt{2 \pi \alpha}} \nonumber \\
&\times \exp \left( \im \frac{p}{\sqrt{2}} \sum_{j=1}^N
\left\{
X_{\nu j} \Theta_j (x) + p (KX)_{\nu j} \Phi_j (x)
\right\}
\right),
\label{eqn:eop} 
\end{align}
where $\eta_\nu$ expresses the Majorana Fermion operator satisfying 
$\eta^\dagger_\nu = \eta_\nu$ and $\left\{\eta_\nu,\eta_{\nu'} \right\} =
2 \delta_{\nu \nu'}$. 
The matrix $X$ is defined as $X = (\bm{X}_1, \bm{X}_2, \cdots,
\bm{X}_N)$.

\subsection{Local density of states}

The LDOS of the $\nu$-th 
band  at the location of bulk $D_{\nu}^{(b)} (\omega,T)$ 
and that of end $D_{\nu}^{(e)} (\omega,T)$ 
are expressed by using these quantities, 
\begin{align}
  & D_{\nu}^{(b/e)} (\omega,T) = \frac{2}{\pi^2 \alpha} 
\prod_{j=1}^N  \left(\frac{\alpha}{v_j}\right)^{Y_{\nu,j}^{(b/e)}}
\nonumber \\
& \times 
\cos \left(\frac{\pi}{2}
 \sum_{j=1}^N Y_{\nu,j}^{(b/e)}\right) 
\int_0^\infty \frac{\d t}{t^{\sum_{j=1}^N Y_{\nu,j}^{(b/e)}}} \nonumber \\
&\times
\left\{
\cos \omega t \left( \frac{\pi Tt}{\sinh \pi Tt} \right)^{\sum_{j=1}^N
 Y_{\nu,j}^{(b/e)}} -1
\right\}. \label{eq_23temp}
\end{align}  
At $T=0$,  the quantities $D_{\nu}^{(b)} (\omega,0)$ and $D_{\nu}^{(e)}
(\omega,0)$ are obtained as follows, 
\begin{align}
  D_{\nu}^{(b/e)} (\omega,0) =& \frac{1}{\pi \alpha}
\prod_{j=1}^N \left( \frac{\alpha}{\tilde{v}_j}\right)^{Y_{\nu,j}^{(b/e)}}
 \nonumber \\
&\times \frac{1}{\Gamma \left[ \sum_{j=1}^N Y_{\nu,j}^{(b/e)}\right]} \omega^{\sum_{j=1}^N
 Y_{\nu,j}^{(b/e)}-1}, \label{eqn:DOS-Omega}
\end{align} 
where $\Gamma[z]$ is the gamma function. 
On the other hand,
the LDOS of the $\nu$-th band at the Fermi energy, 
$D_{\nu}^{(b/e)} (0,T)$ at the bulk/edge is written as 
\begin{eqnarray}
& &
D_{\nu}^{(b/e)} (0,T) = \frac{1}{\pi^2 \alpha}
\prod_{j=1}^N \left( \frac{\alpha}{\tilde{v}_j}\right)^{Y_{\nu,j}^{(b/e)}} \nonumber \\
& &
\times \frac{ \left\{\Gamma \left[\sum_{j=1}^N Y_{\nu,j}^{(b/e)}/2 \right] \right\}^2}
{\Gamma\left[\sum_{j=1}^N Y_{\nu,j}^{(b/e)}\right]}
(2 \pi T)^{\sum_{j=1}^N Y_{\nu,j}^{(b/e)}-1}. \label{eqn:DOS-T} 
\end{eqnarray}
In eqs. (\ref{eqn:DOS-Omega}) and (\ref{eqn:DOS-T}),  
\begin{align}
 Y_{\nu,j}^{(b)} =& \frac12 \left\{ \frac{(X_{\nu,j})^2}{2 \pi \tilde{v}_j} +
 2 \pi \tilde{v}_j \left[(KX)_{\nu,j}
 \right]^2\right\}, \\ 
Y_{\nu,j}^{(e)} =&
 2 \pi \tilde{v}_j \left[(KX)_{\nu,j}\right]^2.
\end{align}
The exponent of the power-law dependence associated with the
$\nu$-th band is expressed by 
\begin{align}
 \lambda^{(b/e)}(\nu) = \sum_{j=1}^N
Y_{\nu,j}^{(b/e)}-1,
\label{eqn:exponent}
\end{align}
and 
the total LDOS, which will be observed experimentally, is given by the sum of contribution from
each band, 
\begin{align}
 D^{(b/e)}(\omega) &= \sum_{\nu=1}^N D_{\nu}^{(b/e)}(\omega,0), \\
 D^{(b/e)}(T) &= \sum_{\nu=1}^N D_{\nu}^{(b/e)}(0,T).   
\end{align}

\section{Power-law exponent}

\subsection{$\lambda^{(b/e)}(\nu)$ for band structure I}

For the band structure I shown in Fig.~\ref{fig:band} (a),
eq.~(\ref{eq_shima25}) leads to   
the velocities of excitation spectra together with the normalized eigenvectors
corresponding to them as 
$\tilde{v}_1 = v_1$ with $\bm{X}_1 = \sqrt{\pi
v_1}(1,0,-1,0)^\mathrm{T}$, 
$\tilde{v}_2 = v_2$ with $\bm{X}_2 = \sqrt{\pi
v_2}(0,1,0,-1)^\mathrm{T}$,
$\tilde{v}_3 = v_+$ with $\bm{X}_3 = 1/\sqrt{Z_+}
(B_+,A_+,B_+,A_+)^\mathrm{T}$ and 
$\tilde{v}_4 = v_-$ with $\bm{X}_4 = 1/\sqrt{Z_-}
(B_-,A_-,B_-,A_-)^\mathrm{T}$, 
where
\begin{equation}
Z_\pm = \frac{B_\pm^2}{\pi v_1} + \frac{A_\pm^2}{\pi v_2},
\end{equation}
\begin{equation}
A_\pm = \frac{v_\pm^2 - v_1^2}{2 \pi v_1}, \;\; 
B_\pm = \frac{v_\pm^2 - v_2^2}{2 \pi v_2},
\end{equation}
\begin{equation}
v^2_\pm = \frac{v_1^2 + v_2^2 + 4 \pi g (v_1 + v_2) \pm \sqrt{D}}{2},
\end{equation}
\begin{eqnarray}
D &=& (v_1^2 - v_2^2) \left\{v_1^2 - v_2^2 + 8 \pi g (v_1 - v_2) \right\} \nonumber \\
& & +\left\{ 4 \pi g (v_1 + v_2) \right\}^2.
\end{eqnarray}
It should be noted 
that the mode with $\tilde{v}_1 = v_1$ and that with $\tilde{v}_2 = v_2$ 
express the spin
excitation of the band 1 and 2, respectively.  
Meanwhile,  
the two charge excitations have the velocities $v_+$ and $v_-$,
respectively.

The above expressions lead to the relation, 
$D_{1}^{(b/e)} (\omega,T) = D_{3}^{(b/e)} (\omega,T)$ and
$D_{2}^{(b/e)} (\omega,T) = D_{4}^{(b/e)} (\omega,T)$, because of  
$Y^{(b/e)}_{1,j}=Y^{(b/e)}_{3,j}$ and $Y^{(b/e)}_{2,j}=Y^{(b/e)}_{4,j}$
($j=1,2,3,4$). 
It reflects the fact that 
the DOS of the 1D band is proportional to the inverse of it's Fermi velocity 
and 
$v_\mathrm{F1} = v_\mathrm{F3} = v_1$ and
$v_\mathrm{F2} = v_\mathrm{F4} = v_2$ in the present case.     
Especially, the exponent of the power-law dependence associate with the
$\nu$-th band, $\lambda^{(b/e)}(\nu)$ given by eq.(\ref{eqn:exponent}) 
are obtained as follows, 
\begin{widetext}
\begin{eqnarray}
\lambda^{(b)}(1) &=& \lambda^{(b)}(3) = 
\frac12 \left\{ 
  \frac{1}{2 \pi v_+} \frac{B_+^2}{Z_+}\left[ 1 + \left( \frac{v_+}{v_1}\right)^2\right]
+ \frac{1}{2 \pi v_-} \frac{B_-^2}{Z_-}\left[ 1 + \left( \frac{v_-}{v_1}\right)^2\right]
\right\}
- \frac12,\\
\lambda^{(b)}(2) &=& \lambda^{(b)}(4) = 
\frac12 \left\{ 
  \frac{1}{2 \pi v_+} \frac{A_+^2}{Z_+}\left[ 1 + \left( \frac{v_+}{v_2}\right)^2\right] 
+ \frac{1}{2 \pi v_-} \frac{A_-^2}{Z_-}\left[ 1 + \left( \frac{v_-}{v_2}\right)^2\right]
\right\}
- \frac12,
\end{eqnarray}
\end{widetext}
\begin{equation}
\lambda^{(e)}(1) = \lambda^{(e)}(3) = 
\frac{v_+}{2 \pi v_1^2}\frac{B_+^2}{Z_+} + \frac{v_-}{2 \pi v_1^2}\frac{B_-^2}{Z_-} - \frac12,
\end{equation}
\begin{equation}
\lambda^{(e)}(2) = \lambda^{(e)}(4) = 
\frac{v_+}{2 \pi v_2^2}\frac{A_+^2}{Z_+} + \frac{v_-}{2 \pi v_2^2}\frac{A_-^2}{Z_-} - \frac12.   
\end{equation}

\subsection{$\lambda^{(b/e)}(\nu)$ for band structure II}

Next, we move to the results for the band structure II in
Fig.\ref{fig:band} (b). 
The velocities of the excitation spectra and the eigenvectors
corresponding to them are analytically obtained as, 
$\tilde{v}_1 = v_1$ with $\bm{X}_1 = \sqrt{\pi v_1}
(1,0,0,-1,0,0)^\mathrm{T}$,
$\tilde{v}_2 = v_2$ with $\bm{X}_2 = \sqrt{\pi v_2}
(0,1,0,0,-1,0)^\mathrm{T}$, 
$\tilde{v}_3 = v_2$ with $\bm{X}_3 = \sqrt{\pi v_2}
(0,0,1,0,0,-1)^\mathrm{T}$,  
$\tilde{v}_4 = v_2$ with $\bm{X}_4 = \sqrt{\pi v_2}
(0,1/\sqrt{2},-1/\sqrt{2},0,1/\sqrt{2},-1/\sqrt{2})^\mathrm{T}$,
$\tilde{v}_5 = v_+$ with $\bm{X}_5 = 1/\sqrt{Z_+}
(B_+,A_+,A_+,B_+,A_+,A_+)^\mathrm{T}$, and 
$\tilde{v}_6 = v_-$ with $\bm{X}_6 = 1/\sqrt{Z_-}
(B_-,A_-,A_-,B_-,A_-,A_-)^\mathrm{T}$.
Here, 
\begin{equation}
Z_\pm = \frac{B_\pm^2}{\pi v_1} + \frac{2 A_\pm^2}{\pi v_2},
\end{equation}
\begin{equation}
A_\pm = \frac{v_\pm^2 - v_1^2}{2 \pi v_1}, \;\;
B_\pm = \frac{v_\pm^2 - v_2^2}{2 \pi v_2},
\end{equation}
\begin{equation}
v_\pm^2 = \frac{v_1^2 + v_2^2 + 4 \pi g (v_1 + 2 v_2) \pm \sqrt{D}}{2},
\end{equation}
\begin{eqnarray}
D &=& \left( v_1^2 - v_2^2 \right) 
\left\{v_1^2 - v_2^2 + 8 \pi g (v_1 - 2 v_2) \right\} \nonumber \\
& &+ \left\{ 4 \pi g (v_1 + 2 v_2)\right\}^2.
\end{eqnarray}
The eigenvectors indicate that
the first, second and third mode express the spin excitation in the band
1, 2 and 3, respectively.
On the other hand, the fourth, fifth and sixth mode correspond to the charge
fluctuation. 
Especially, the fourth mode expresses antisymmetric charge excitation
between the band $2$ and $3$.

Similar to the case I,
the following relations hold:
$D_{1}^{(b/e)} (\omega,T) = D_{4}^{(b/e)} (\omega,T)$ and
$D_{2}^{(b/e)} (\omega,T) = D_{3}^{(b/e)} (\omega,T) = D_{5}^{(b/e)} (\omega,T) = D_{6}^{(b/e)} (\omega,T)$.
Especially, the exponents of power-law dependences are obtained as follows.
\begin{widetext}
\begin{eqnarray}
\lambda^{(b)}(1) &=& \lambda^{(b)}(4) = 
\frac12 \left\{ 
  \frac{1}{2 \pi v_+} \frac{B_+^2}{Z_+}\left[ 1 + \left( \frac{v_+}{v_1}\right)^2\right]
+ \frac{1}{2 \pi v_-} \frac{B_-^2}{Z_-}\left[ 1 + \left( \frac{v_-}{v_1}\right)^2\right]
\right\} - \frac12,\\
\lambda^{(b)}(2) &=& \lambda^{(b)}(3) = \lambda^{(b)}(5) =  \lambda^{(b)}(6) 
= \frac12 
\left\{ 
\frac{1}{2 \pi v_+} \frac{A_+^2}{Z_+}\left[ 1 + \left( \frac{v_+}{v_2}\right)^2\right]
+
\frac{1}{2 \pi v_-} \frac{A_-^2}{Z_-}\left[ 1 + \left( \frac{v_-}{v_2}\right)^2\right]
\right\}
- \frac{1}{4},
\end{eqnarray}
\end{widetext}
\begin{eqnarray}
\lambda^{(e)}(1) &=& \lambda^{(e)}(4) =
\frac{v_+}{2 \pi v_1^2}\frac{B_+^2}{Z_+} + \frac{v_-}{2 \pi v_1^2}\frac{B_-^2}{Z_-} - \frac12, \\
\lambda^{(e)}(2) &=& \lambda^{(e)}(3) = \lambda^{(e)}(5) = \lambda^{(e)}(6) \nonumber \\
&=& \frac{v_+}{2 \pi v_2^2}\frac{A_+^2}{Z_+} + \frac{v_-}{2 \pi v_2^2}\frac{A_-^2}{Z_-} - \frac{1}{4}.
\end{eqnarray}

\section{Results and Discussion}

\subsection{Numerical evaluation of $\lambda^{(b/e)}(\nu)$}

\begin{figure}[ttt]
 \begin{center}
  \includegraphics[height=12truecm,clip]{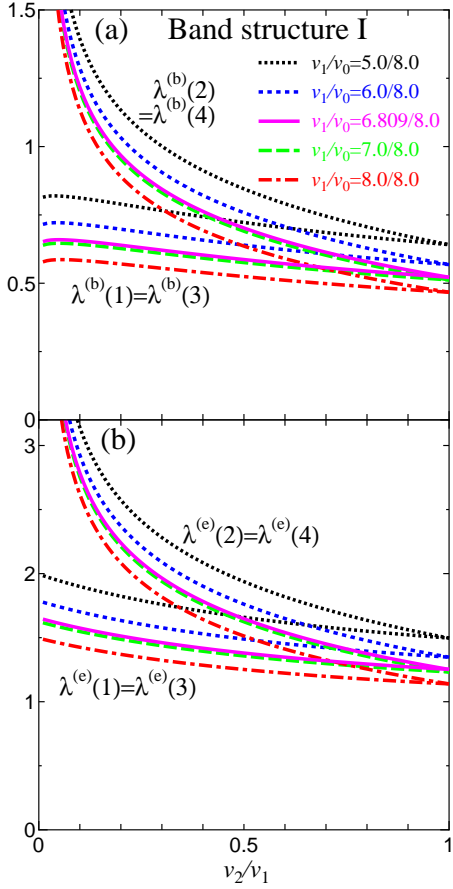}
 \end{center}
\caption{(a) Exponents in LDOS at the bulk, $\lambda^{(b)}(\nu)$, 
as a function of the ratio of the Fermi velocities $v_2/v_1$ 
for the band structure I.
Here, $\nu$ ($\nu = 1,2,3,4$) expresses the band index.
(b) Counterparts at the edge, $\lambda^{(e)}(\nu)$, for the band structure I.
}
\label{fig:dos_exponent}
\end{figure}
\begin{figure}[ttt]
 \begin{center}
  \includegraphics[height=12truecm,clip]{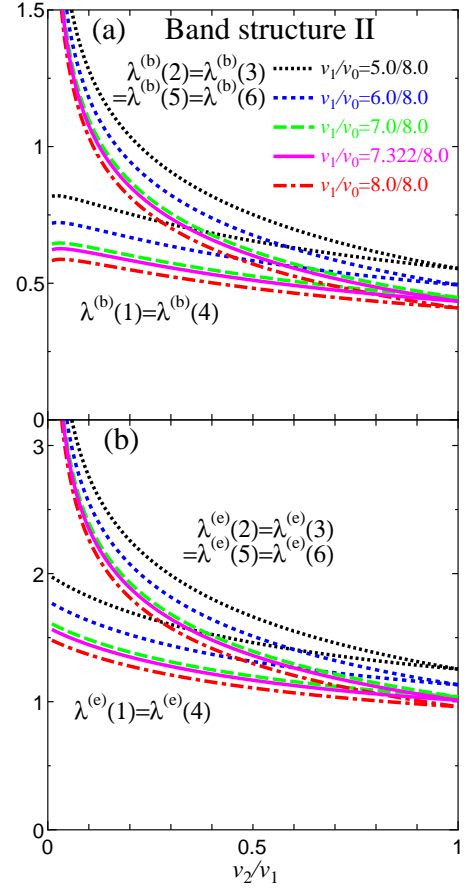}
 \end{center}
\caption{
(a) Bulk-exponents in LDOS, $\lambda^{(b)}(\nu)$ with $\nu = 1,2,3,4,5,6$, 
as a function of $v_2/v_1$ for the band structure II.
(b) Edge-exponents in LDOS, $\lambda^{(e)}(\nu)$, for the band structure II.
}
\label{fig:dos_exponent_MS}
\end{figure}

Figure \ref{fig:dos_exponent} demonstrates
the exponents for the band structure I.
The bulk exponent $\lambda^{b}(\nu)$ and the edge exponent $\lambda^{e}(\nu)$
are plotted in Figs.~\ref{fig:dos_exponent}(a) and \ref{fig:dos_exponent}(b), respectively,
as a function of $v_2/v_1$ for several choices of $v_1/v_0$.
The interaction strength was set to be $\bar{V}(0)/(\pi v_0) = 7.47$,
so that $\lambda^{b}(\nu)$ should be equal to the bulk exponent of the
carbon nanotube \cite{Ishii2003}
under the condition of $v_1 = v_2 = v_0$. 
The bulk exponent of the carbon nanotubes 
is indicated by the intersection of the dashed-dotted curves (colored in red)
with $v_1/v_0 = 1$ with the vertical line with $v_2/v_1 = 1$
in Fig.~\ref{fig:dos_exponent} (a).
It is clearly seen that both $\lambda^{(b)}(1)$ and $\lambda^{(b)}(2)$
for the band structure I
exceed the bulk exponent of carbon nanotubes,
regardless of the choices of the parameters $v_1$ and $v_2$.

It follows from Fig.~\ref{fig:dos_exponent} that
the band with a larger $v_1$ leads to a smaller $\lambda^{(b/e)}(1)$.
It is because the electronic correlation is effectively suppressed 
with increasing the Fermi velocity.\cite{Yoshioka2011}
Furthermore, for a given $v_1$,
$\lambda^{(b/e)}(1)$ falls behind $\lambda^{(b/e)}(2)$ at any $v_2$.
Our numerical data of $v_1$ and $v_2$ for the FP6L model (see Table \ref{table_fermi})
allow to estimate the exponents for the band structure I as:
\begin{eqnarray}
& &\lambda^{b}(1) = 0.56844, \;\; \lambda^{b}(2) = 0.64506, \nonumber \\
& &\lambda^{e}(1) = 1.3416, \;\; \lambda^{e}(2) = 1.5211.
\label{eq_lambda_11}
\end{eqnarray}
Similar trends in the $v_1$- and $v_2$-dependences of $\lambda^{(b/e)}(\nu)$
have been confirmed for the band structure II, too,
as seen from Fig.~\ref{fig:dos_exponent_MS}.
The exponents for the band structure II can be evaluated as
\begin{eqnarray}
& &\lambda^{b}(1) = 0.54178, \;\; \lambda^{b}(2) = 0.68746, \nonumber \\
& &\lambda^{e}(1) = 1.2388, \;\; \lambda^{e}(2) = 1.5413.
\label{eq_lambda_22}
\end{eqnarray}

It should be reminded that 
the smallest $\lambda^{b/e}(\nu)$ 
({\it i.e.}, the contribution from the largest-$v_F(\nu)$ energy band) 
takes a primary role in experimental observations of low-energy (low-temperature)
physics of TLL states.
This implies that the power-law anomalies of the 1D \fu polymers
are governed by the exponent $\lambda^{(b/e)}(1)$
particularly at low energies (temperatures),
regardless of which model, FP6L or FP5N, describes the materials
actually synthesized in the experiments.
In contrast, at relatively high energy (temperature) range,
the other exponent $\lambda^{(b/e)}(2)$ may contribute to the power-law anomalies.

\begin{figure}[ttt]
 \begin{center}
  \includegraphics[height=6truecm,clip]{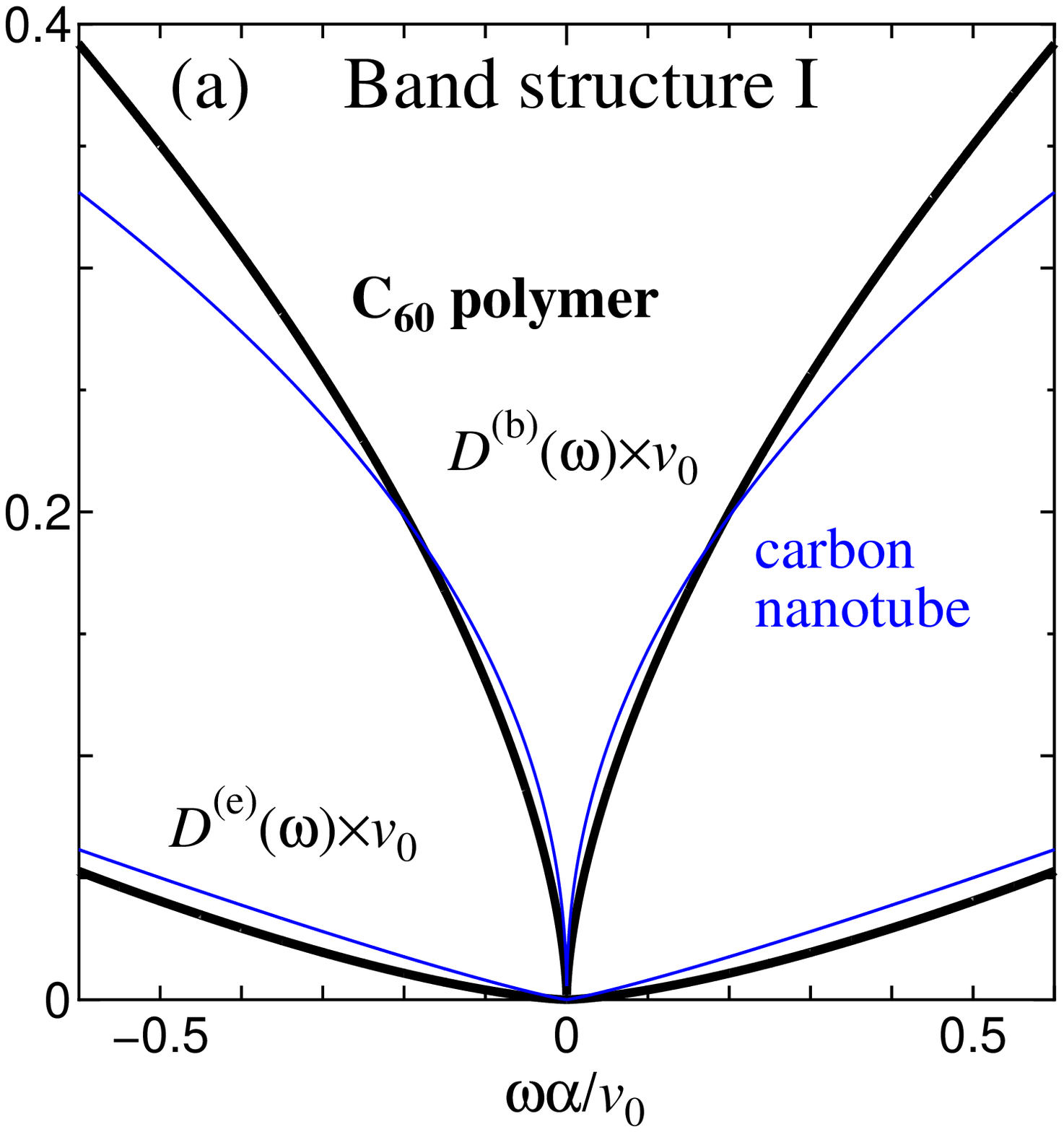}
  \includegraphics[height=6truecm,clip]{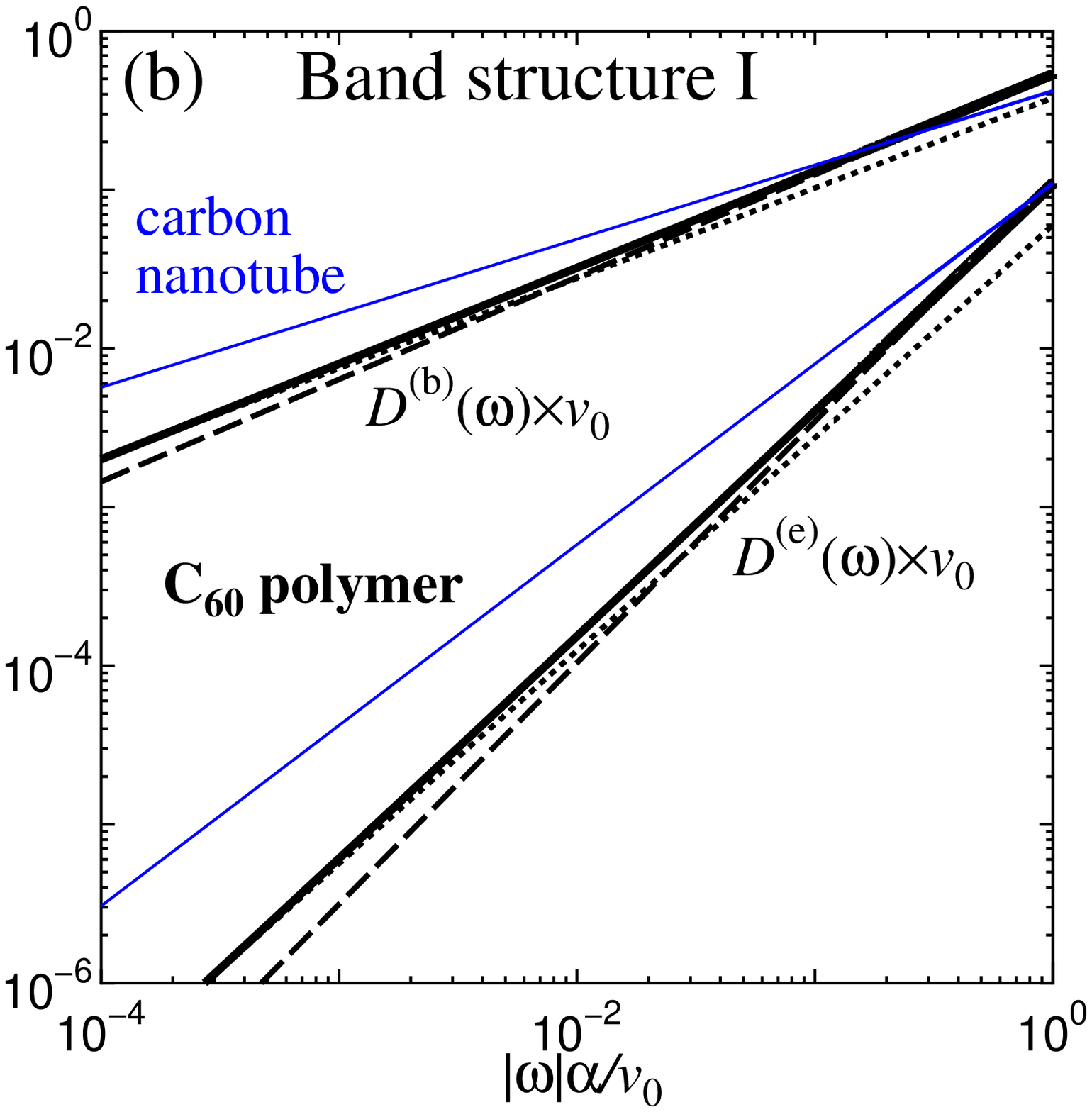}  
 \end{center}
\caption{Normalized LDOS  for the band structure I at the bulk position
 $D^{(b)}(\omega) \times v_0$
 and at the edge $D^{(e)}(\omega) \times v_0$ as a function of the frequency
 $\omega$ divided by $v_0/\alpha$ with $v_0$ and $\alpha$ the Fermi velocity of
 the graphene sheet and the short distance cut-off. 
The linear and double logarithmic plot are shown in (a) and (b),
 respectively.
Thick solid, thick dashed, and thick dotted curves (all colored in black)
express the quantities of \fu polymer with the
 Fermi velocities given in Table \ref{table_fermi}. 
Thin solid curves indicate the counterparts for the carbon nanotubes.
}
\label{fig:dos_w-dep_1}
\end{figure}
\begin{figure}[ttt]
 \begin{center}
  \includegraphics[height=6truecm,clip]{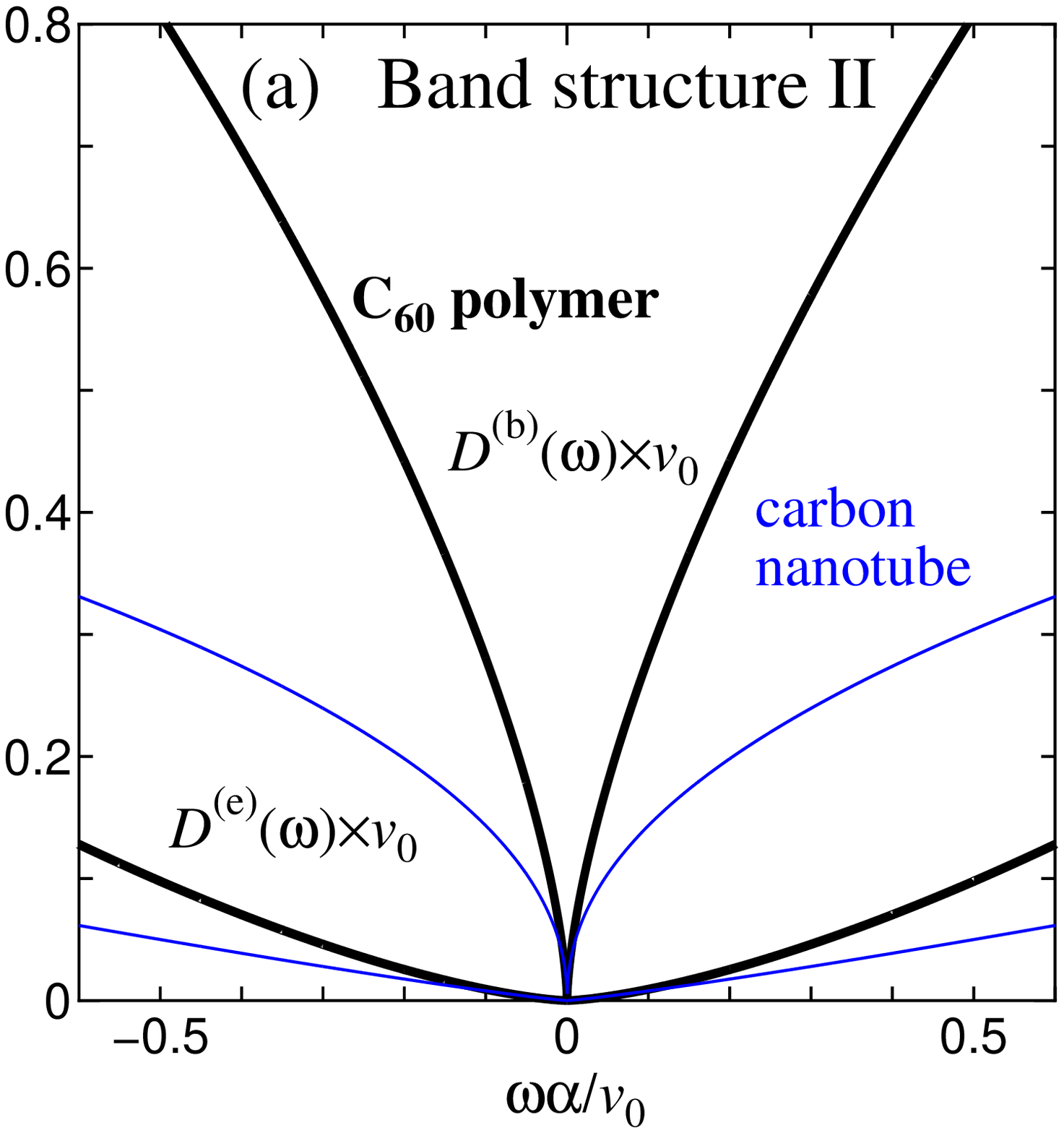}
  \includegraphics[height=6truecm,clip]{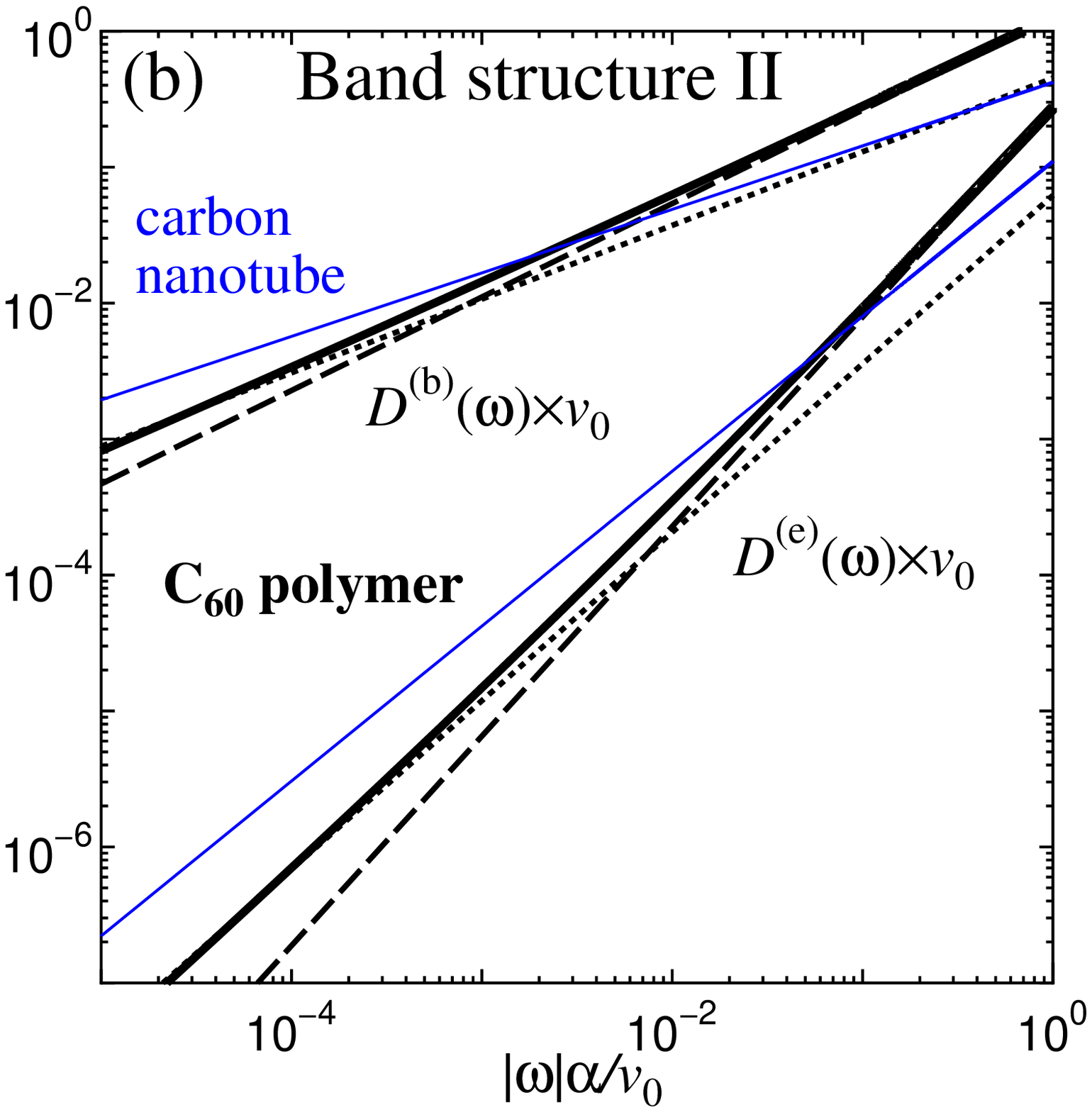}  
 \end{center}
\caption{The $\omega$-dependence of the normalized LDOS for the band structure II.
The style for plot is same as that in Fig.~\ref{fig:dos_w-dep_1}.
}
\label{fig:dos_w-dep_2}
\end{figure}
\begin{figure}[ttt]
 \begin{center}
  \includegraphics[height=6truecm,clip]{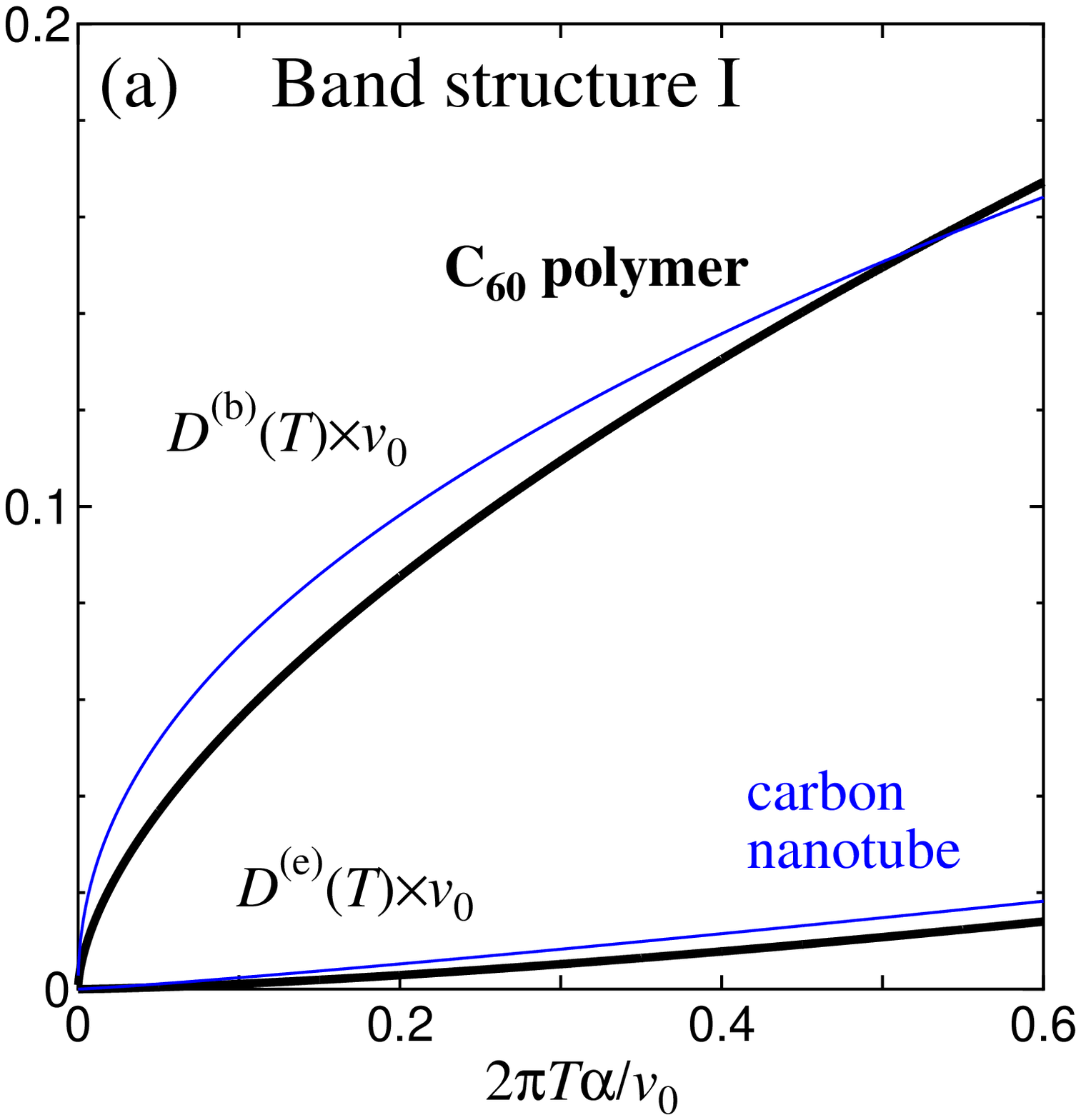}
  \includegraphics[height=6truecm,clip]{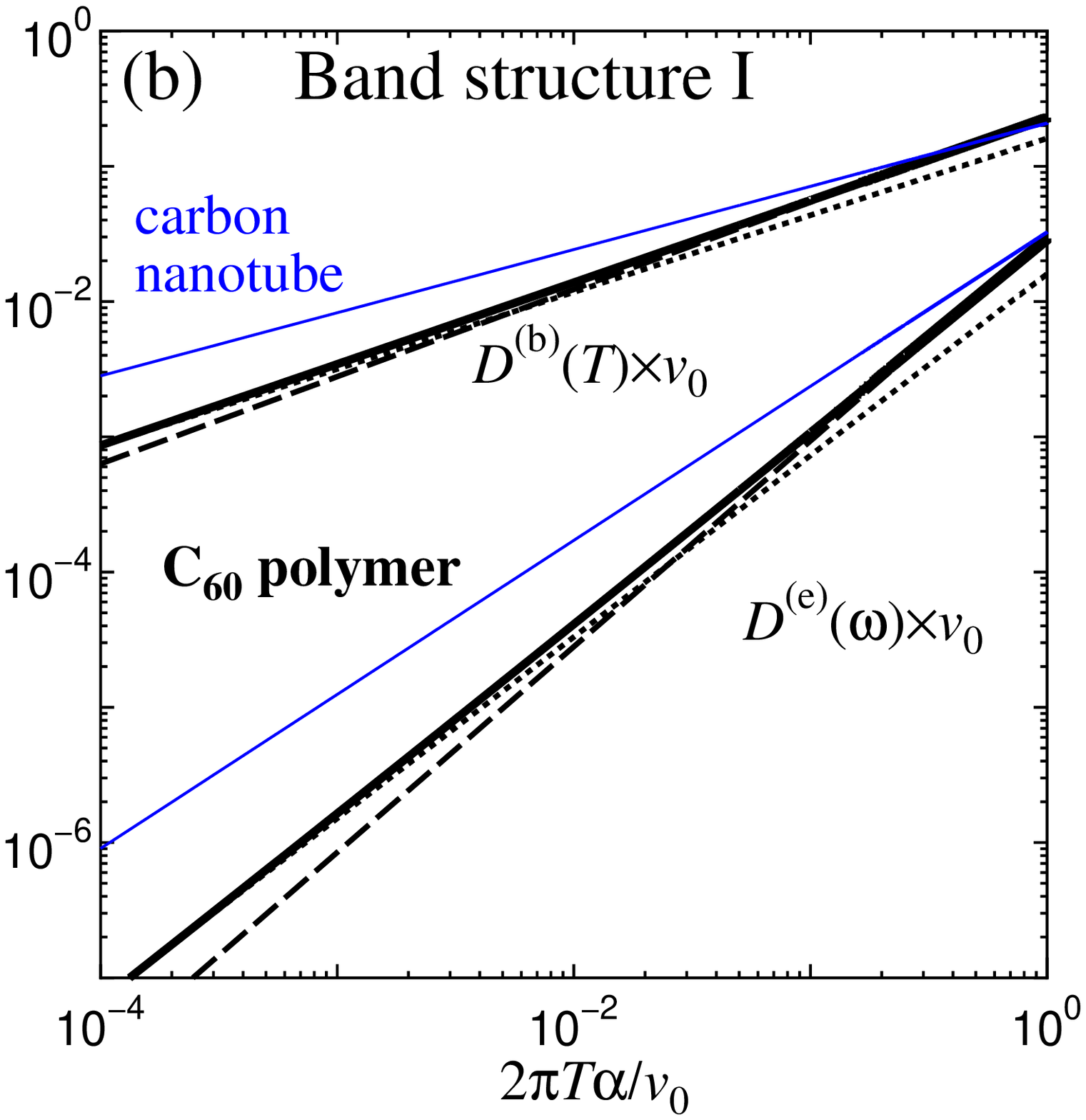}  
 \end{center}
\caption{The $T$-dependence of the normalized LDOS for the band structure I:
$D^{(b)}(T) \times v_0$ and $D^{(e)}(T) \times v_0$ as a function of the temperature
 $T$ divided by $v_0/(2 \pi \alpha)$.
}
\label{fig:dos_T-dep_1}
\end{figure}
\begin{figure}[ttt]
 \begin{center}
  \includegraphics[height=6truecm,clip]{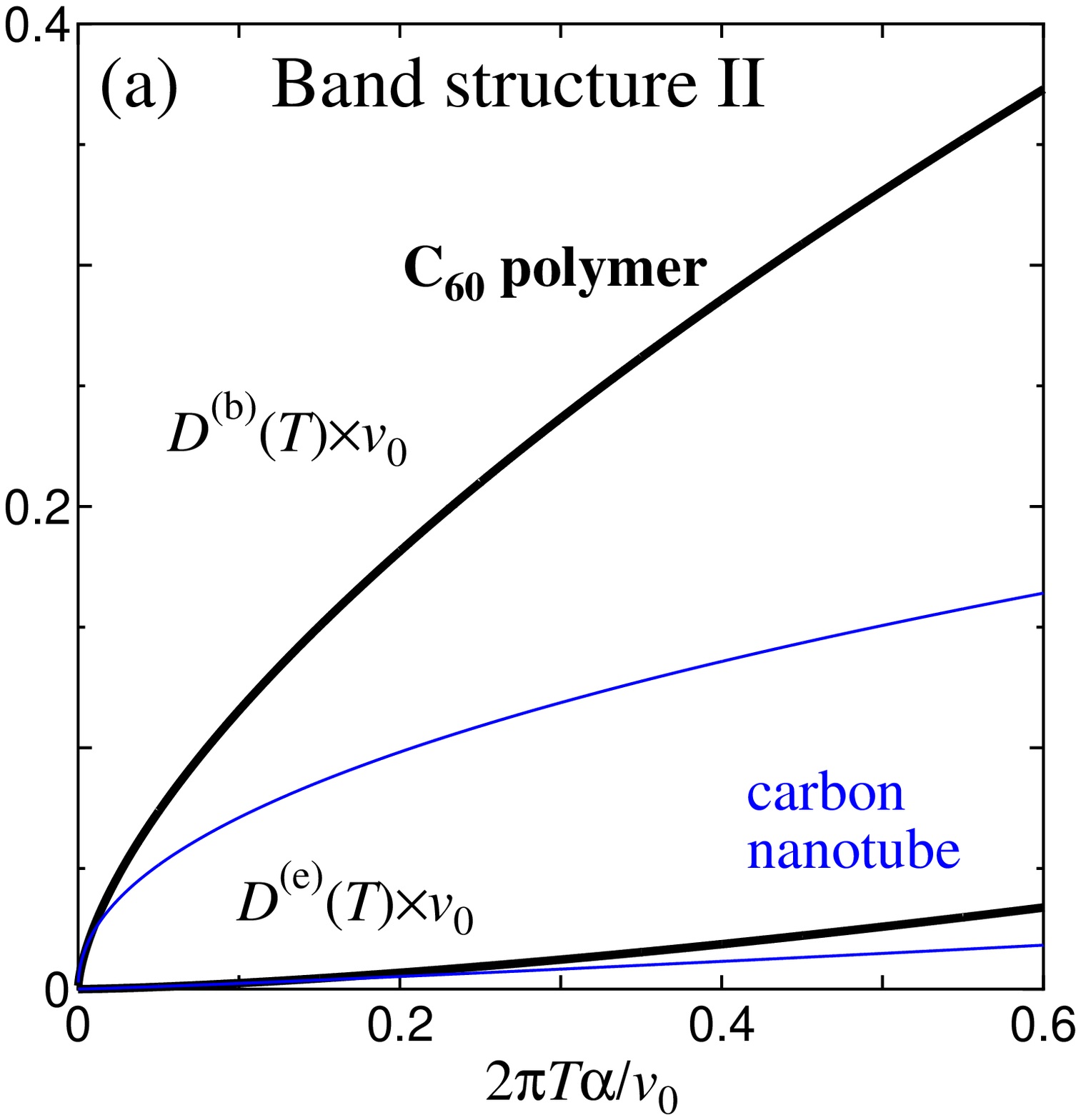}
  \includegraphics[height=6truecm,clip]{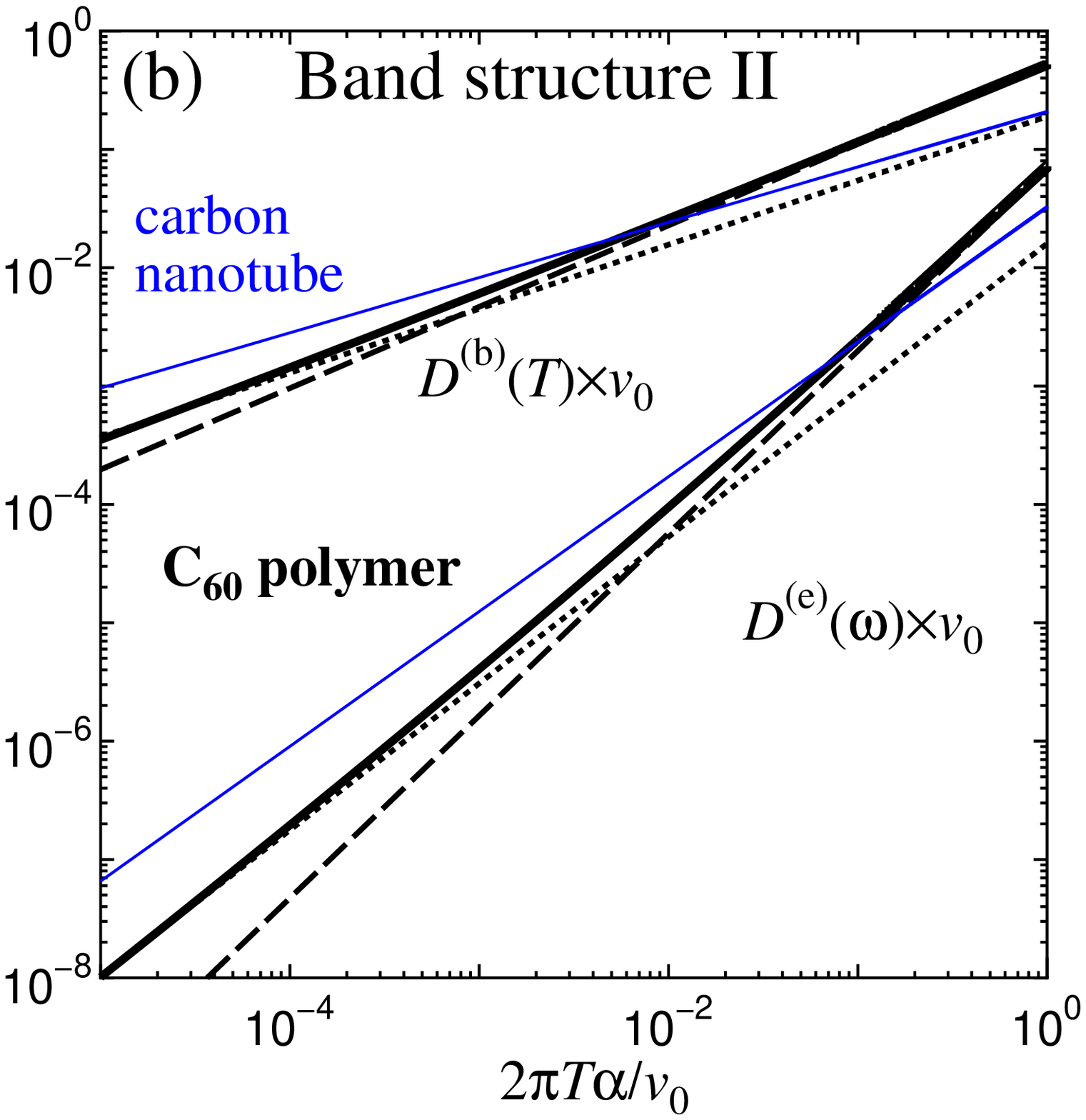}  
 \end{center}
\caption{The $T$-dependent normalized LDOS for the band structure II.
}
\label{fig:dos_T-dep_2}
\end{figure}

\subsection{Crossover in the power-law of LDOS}

The total LDOS is given by the sum of contribution from
each band, 
\begin{align}
 D^{(b/e)}(\omega) &= \sum_{\nu=1}^N D_{\nu}^{(b/e)}(\omega,0), \\
 D^{(b/e)}(T) &= \sum_{\nu=1}^N D_{\nu}^{(b/e)}(0,T),   
\end{align} 
where $N=4$ or $N=6$ for the band structure I or II, respectively.
Figures \ref{fig:dos_w-dep_1} and \ref{fig:dos_w-dep_2} show 
the $\omega$-dependence of $v_0 D^{(b/e)}(\omega)$,
{\it i.e.,} the total LDOS normalized by $v_0^{-1}$:
Fig.~\ref{fig:dos_w-dep_1} corresponds to the band structure I,
while Fig.~\ref{fig:dos_w-dep_2} the band structure II.
The panels (a) and (b) display
the linear and double logarithmic plots, respectively.
In all panels, counterpart results for the carbon nanotubes are displayed 
by the thin solid curves.

Our numerical results clearly demonstrate that 
the total LDOS of the 1D \fu polymer close to Fermi energy $\omega \simeq 0$ is suppressed compared to
that of carbon nanotubes. 
It means that the electronic correlation is effectively stronger in the
1D \fu polymer because of the smaller Fermi velocity.
In Figs. \ref{fig:dos_w-dep_1} (b) and \ref{fig:dos_w-dep_2} (b), 
the dotted and dashed line express the fitting by 
$A \omega^{\lambda^{(b/e)}(1)}$ and $B \omega^{\lambda^{(b/e)}(2)}$,
respectively. 
For the low energy region, 
the components with the smaller exponent is dominant in the LDOS. 
On the other hand, the major part is given by those with the larger exponent
for the higher energy part.
Namely, we have
\begin{equation}
D^{(b/e)}(\omega) \propto 
\begin{cases}
\omega^{\lambda^{(b/e)}(1)} \;\; \mbox{at} \;\;
|\omega|\ll c v_0/\alpha,\\ 
\omega^{\lambda^{(b/e)}(2)} \;\; \mbox{at} \;\;
|\omega|\gg c v_0/\alpha
\end{cases}
\label{eq_crossover_omega}
\end{equation}
The constant $c$ in eq.~(\ref{eq_crossover_omega})
depends on both the position of contact (bulk or edge)
and the band structure (I or II).
The constant lies inbetween $10^{-3}$ and $3\times 10^{-2}$
at a rough estimate from the figures.
Provided $v_0 = 8.0 \times 10^5$ m/s and $\alpha=3.5$ \AA, therefore,
the 1D \fu polymer is expected to show an crossover at an energy on the order of 100 meV more or less.
Such the crossover has not yet been observed in experiments of the 1D \fu polymers,
nor suggested by earlier theoretical work.
The expected variations in the exponents across the crossover
are summarized in Table \ref{table_exponentnew}.

It should be remarked that the crossover behavior in the power law LDOS
is a direct consequence of the multichannel conduction mechanism
with different Fermi velocities.
Therefore, our prediction on the crossover occurrence may apply to
1D quantum nanomaterials other than \fu polymers,
as far as electronic dispersion curves near the Fermi energy
constitute a slanted band crossing as similar to those in Fig.~\ref{fig:band}.

\begin{table}[bbb]
\caption{Power-law exponents at frequencies lower and higher than the crossover frequency $\omega_c$.}
\begin{tabular}{|c|c|c|c|c|}
\hline
 & \multicolumn{2}{c|}{Band structure I} & \multicolumn{2}{c|}{Band structure II} \\ \hline
 & low-$\omega$ & high-$\omega$ & low-$\omega$ & high-$\omega$  \\ \hline
\;\; Bulk \;\; & \;\; $0.57$ \;\; & \;\;$0.65$ \;\; & \;\; $0.54$ \;\; & \;\;$0.69$ \;\; \\
\;\; Edge \;\; & \;\; $1.3$ \;\; & \;\;$1.5$ \;\; & \;\; $1.2$ \;\; & \;\;$1.5$ \;\; \\ \hline
\end{tabular} 
\label{table_exponentnew}
\end{table}

Two important consequences are drawn from eq.~(\ref{eq_crossover_omega}).
First, our theoretical result on the low-energy power-law anomaly is in good agreement
with the experimental finding based on the high-resolution PES measurement.\cite{Onoe2012}
It was concluded in Ref.~\onlinecite{Onoe2012}
that the DOS for the 1D \fu polymes obeys
$D(\omega)\propto \omega^\lambda$ with $\lambda=0.65\pm 0.08$
at 18-100 meV.
The exponent was found to be significantly larger than that of $0.43$-$0.54$
for metallic carbon nanotubes.
The multichannel TLL theory we have developed in the present work
describes successfully the physical origin of the overshoot
in the power-law exponent of DOS in a quantitative manner.
Second, eq.~(\ref{eq_crossover_omega}) indicates the possibility that
edge-contact measurements will give an exponent greater than one,
as implied by eqs.~(\ref{eq_lambda_11}) and (\ref{eq_lambda_22}).
It is hoped that this theoretical consequence can be
verified experimentally in the measurements of
LDOS and/or electronic conductance of the 1D \fu polymers
under the edge-contact condition.

Figures \ref{fig:dos_T-dep_1} and \ref{fig:dos_T-dep_2}
demonstrate the $T$-dependences of the total LDOS, $D^{(b/e)} (T)$,
for the band structures I and II, respectively.
Similarly to the $\omega$-dependent LDOS,
$D^{(b/e)} (T)$ behaves as $A' T^{\lambda^{(b/e)}(1)}$ and $B' T^{\lambda^{(b/e)}(2)}$
at the low and high temperature region, respectively.
The crossover temperature $T_c$ is estimated to be on the order of $10^3$ K,
across which the power-law exponent switches.
Again, the $T$-dependence of the LDOS at temperatures lower than $T_c$
agrees quantitatively with the experiment reported in Ref.~\onlinecite{Onoe2012}.

\section{Summary}

In the present contribution, we have examined the LDOS of the 1D \fu polymer,
a kind of multi-channel 1D systems involving different Fermi velocities.
The Fermi velocities of the 1D \fu polymer have been determined 
using the first principles calculation
that enabled us to evaluate the energetically stable atomic configuration
and the resulting electronic band structures.
Through a multichannel bosonization approach,
we have established a closed-form solution for the power-law dependent LDOS.
The solution is in a good agreement with the experimental finding
obtained by the earlier PES measurement.
Furthermore, 
the solution suggests that different Fermi-level-crossing bands give different power-law dependent LDOS,
implying the occurrence of a crossover in the LDOS spectrum at energy on the order of 100 meV.
It is inferred that the predicted crossover may be observed various 1D systems
as well as the 1D \fu polymers,
as far as they are endowed with multichannel and slanted band crossing at the Fermi level.
We hope that the present results would inspire experimental efforts
to verify the predicted crossover and elucidate the multichannel effects
on the LDOS spectra and transport anomalies of the TLL systems.

\section*{Acknowledgement}
This work was supported by JSPS KAKENHI Grant Numbers 25390147, 25400370, and 15K04619.
Fruitful discussions with Jun Onoe would be greatly appreciated.
Shima is grateful for the financial support from the Noguchi Institute.


%




\end{document}